# I. Introduction

Biliary tract cancers or cholangiocarcinomas (CCA) are a group of biliary tract malignant and aggressive tumor with poor prognosis (Fig. 1) [1]. They represent the second most frequent primary liver cancers and 3% of all gastrointestinal neoplasia. CCA present an overall tendency to increase presenting an incidence between 2-6 cases per 100,000 people [2]. Importantly, this cancer is usually diagnosed in advance stages due to its asymptomatic behaviour and the difficulties in monitoring high-risk patients, especially primary sclerosing cholangitis (PSC) patients [1] [3]. Nowadays surgical resection of the tumor is still the only curative and effective therapeutic alternative. However, due to the advanced stage of the disease at the diagnosis, only low proportion of patients can benefit of this treatment and the recurrence is high [2]. When tumor resection is not possible or at recurrence, the therapeutic alternatives consist in palliative treatments based on chemotherapy regimens with poor results, usually patients undergoing palliative chemotherapy present survivals below a year [2] [4]. Faced with this alarming finding, the emergence of new therapeutic approaches is highly expected. Cold plasmas represent a new research avenue where conventional therapies appear limited.

For more than one year, the LPP and the Saint-Antoine Research Center (CRSA) carry out joint research on the treatment of carcinomas using atmospheric pressure plasma jets (APPJ) as therapeutic option. The objective is to determine an experimental operating window where plasma therapy can be applied first as a safe approach (i.e. without inducing deleterious effects on the tissues) and second as an effective approach (i.e. demonstrating antitumor effects on *in vivo* models).

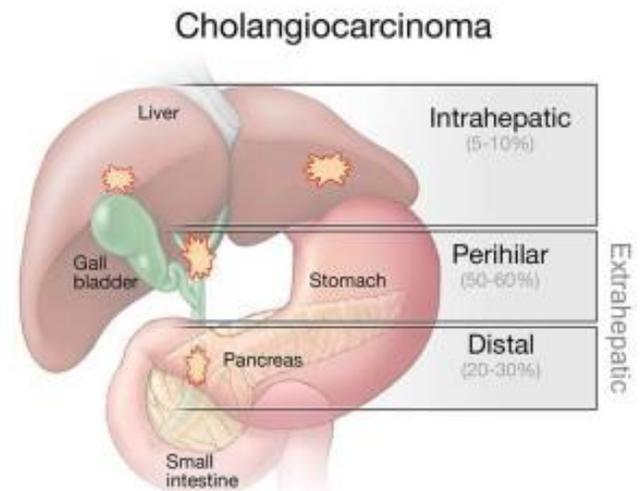

*Fig. 1. Anatomical classification of CCAs, percentages represent the frequency of each CCA subtype.*

# II. Experimental setup

The Fig. 2a shows a configuration of the plasma device engineered at LPP and dedicated to medical applications. Supplied with a high voltage nanoimpulse generator, it can deliver pulsed atmospheric plasma streams (PAPS), i.e. ionization waves with high electric field fronts. As photographed in Fig. 2b, the device can be designed using shorter dielectric tubes as the one utilized at CRSA to treat ectopic CCA tumors on mice models. This "smaller device" has been utilized for *in vivo* campaigns considering two different types of electrode configurations: the "plasma gun" configuration, where plasma can directly interact with the coaxial HV electrode and the "plasma Tesla jet" configuration, where plasma cannot be in contact with the HV electrode.

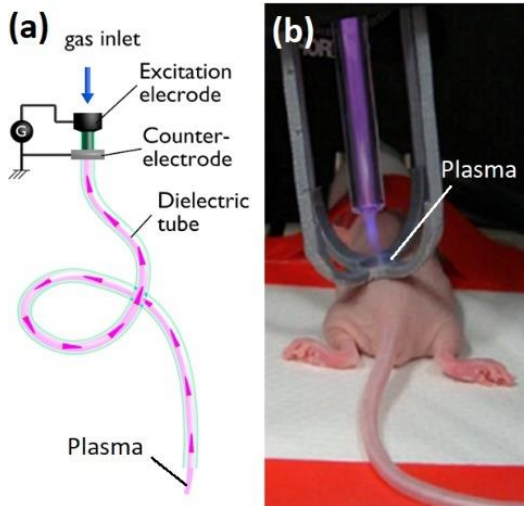

*Fig. 2. (a) Overall view of the APPJ utilized for plasma medicine issues, (b) picture of the APPJ in plasma gun configuration applied on a mouse model to treat carcinoma.*

## III. Calibration and characterization of plasma sources interacting with targets (at physics laboratory)

Before carrying out *in vivo* campaigns at CRSA, physico-chemical properties of the two APPJ configurations have been studied while interacting with different types of targets: liquid interfaces, dielectric and metal plates as well as targets mimicking the electrical properties of living organisms.

The relative efficiency of the APPJ in producing reactive species has been studied while interacting with aqueous interfaces. In the case of the plasma gun configuration, Fig 3a and 3b show how [$NO_2^-$] and [$H_2O_2$] are increased as a function of the duty cycle and repetition frequency parameters. Highest amounts of reactive species at the lowest plasma energy ($E_{PL}$) are obtained for repetition frequency close to 30 kHz and duty cycle lower than 10%. The values of these electrical parameters have been utilized for the *in vivo* campaigns.

Besides, gas flowing dynamics have been studied using space resolved mass spectrometry as shown in Fig. 4 for plasma interacting with dielectric and metal targets as well as an equivalent electrical human body target (EEHB). This latter model includes a resistor of 1500 Ω and a capacitor of 100 pF [5]. For the same operating conditions, the helium and air flow space distributions can reveal quite different and – in some cases – drive to electrical hazards. As an example, the space distributions of helium and air appear monodirectional with metal target at floating potential while it broadens on dielectric targets, especially in the case of the plasma Tesla jet. Besides, while no electrical hazards are detectable using dielectric and metal plates, the utilization of the EEHB target permits to evidence in some cases glow-to-arc transitions, likely to cause severe damages of tissues and endanger patient safety.

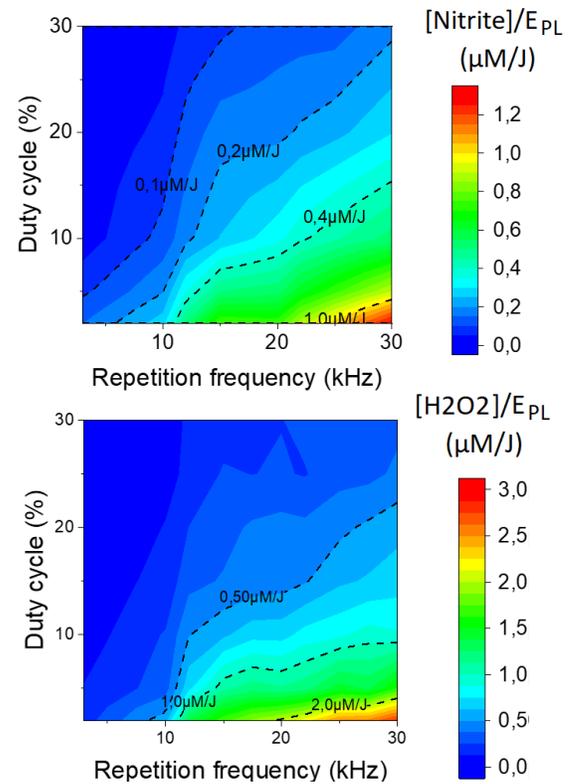

*Fig. 3. Influence of plasma electrical parameters (repetition frequency and duty cycle) on the energetic molar production rates of (a) nitrites and (b) $H_2O_2$, considering the plasma gun configuration.*

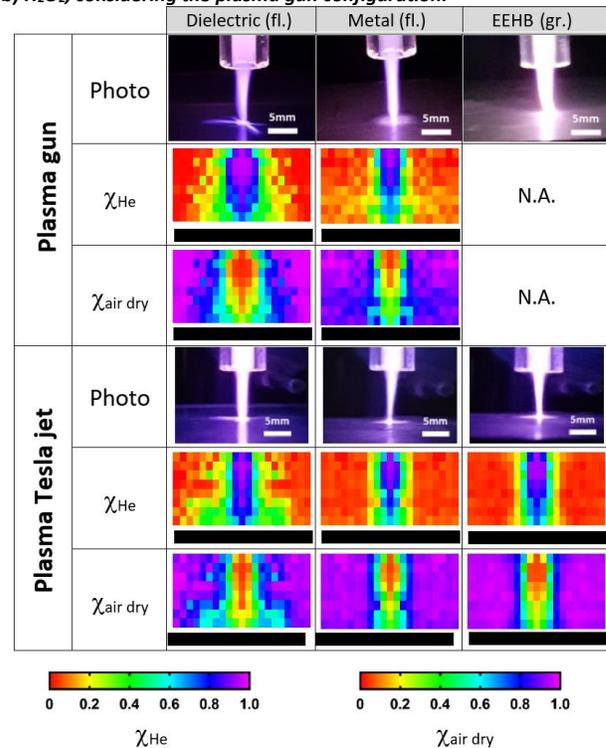

*Fig. 4. Mass spectrometry 2D profiles of plasma plumes generated either with plasma gun or plasma Tesla jet interacting with various targets either at floating potential (dielectric and metal target) or at grounded potential (Equivalent Electrical Human Body target)*

**2**

# IV. Innocuity and therapeutic efficiency (*in vivo* campaigns)

Plasma innocuity has been studied first by treating healthy tissues of mice models and second by assessing skin anatomy and physiology through H&E analysis as well as camera imaging for several days. While mild and severe damages could be obtained in specific conditions using our plasma gun configuration (Fig. 5), no deleterious effects were observed on the tissues exposed to the plasma Tesla jet device. H&E analyses confirmed these observations after examination of hypodermis, dermis and epidermis layers.

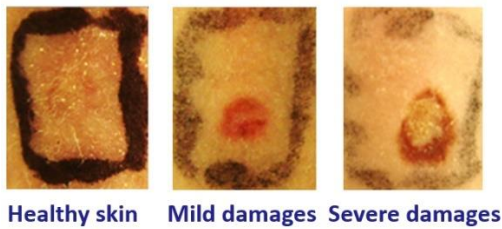

*Fig. 5. Types of healthy skin damages obtained after exposure to APPJ in plasma gun configuration.*

The search for anti-tumor effects was the subject of a first study with the plasma gun configuration. For this, two groups of 8 mice with subcutaneous CCA were carried out: a control group (no treatment) and a plasma group where tumors were exposed to plasma for 2 min/session. The evolution of tumor volumes as a function of time is shown in Fig. 6. After two plasma exposures at days 13 and 20, no antitumor effect was found since no difference in tumor volume could be evidenced between control and plasma groups. Then, as an alternative, the second APPJ device was utilized using same experimental conditions at days 34 and 41. As a result, Fig. 6 clearly shows that the tumor volume rise is significantly reduced in the case of the plasma group [6].

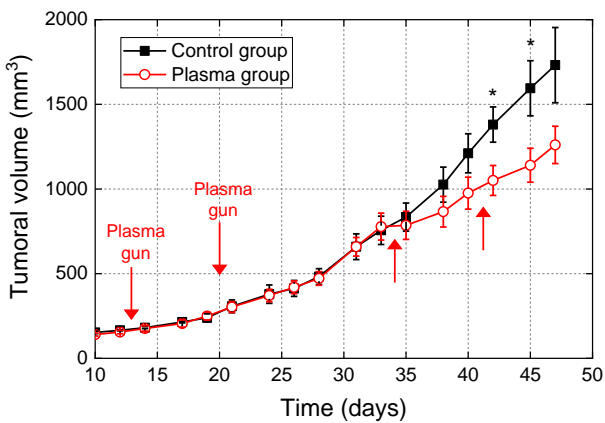

*Fig. 6. Tumor volume follow-up as a function of time considering two mice groups: control (N=8) and plasma (N=8). Plasma treatments are achieved using the plasma gun at days 13 and 20 while plasma Tesla jet at days 34 and 42 [6].*

We have also compared the efficacy of plasma therapy with conventional treatment based on chemotherapy. In Fig. 7, a time follow-up of tumor volumes is represented for three groups of mice with subcutaneous tumors: a control group, a group of mice exposed to plasma and a group of mice treated with gemcitabine (drug administered to patients in palliative care). Red arrows indicate the frequency of plasma or gemcitabine treatments (i.e. approximately every 3 days). It appears that tumor volumes treated with gemcitabine are greatly reduced and stabilize at day 35, while the mean tumor volume of the plasma group is significantly reduced compared to the control group. Tumors extracted from their hosts are photographed on the right of Fig. 3: while the tumors in the control group are the largest and present a vascular system, tumors in the plasma group appear smaller and undamaged.

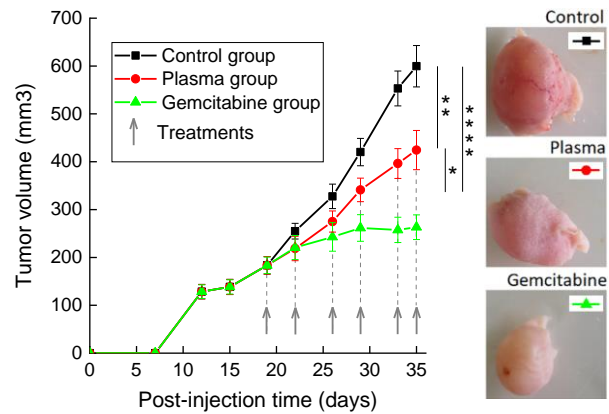

*Fig. 7. Tumor volume follow-up as a function of time considering three mice groups: control (N=8) plasma (N=8) and gemcitabine (N=8).*

# V. References

[1] J. M. Banales, V. Cardinale, G. Carpino, M. Marzioni, et al., Nature Reviews Gastroenterology & Hepatology, 2016, 261-280
[2] S. Rizvi, S. A. Khan, C. L. Hallemeier, R. K. Kelley, G.J. Gores, Nat Rev Clin Oncol., 2018, 95-111
[3] N., Razumilava; G.J., Gores, Cholangiocarcinoman, The Lancet, 2014, 2168-2179
[4] C. Fitzmaurice, D. Dicker, A. Pain, et al., The Global Burden of Cancer 2013, JAMA oncology, 2015, 505-527
[5] F. Judée, T. Dufour, Plasma gun for medical applications: engineering an equivalent electrical target of human body and deciphering relevant electrical parameters , Journal of Physics D: Applied Physics, Letter, 2019, 119262, https://doi.org/10.1088/1361-6463/ab03b8
[6] F. Judée, J. Vaquero, S. Guégan, L. Fouassier, T. Dufour, Atmospheric pressure plasma jets applied to cancerology: correlating electrical configuration with in vivo toxicity and therapeutic efficiency, Submitted to Journal of Physics D: Applied Physics, ref. 119442.